\documentclass[aps,prl,twocolumn,groupedaddress]{revtex4}

\usepackage[pdftex]{color,graphicx}
\usepackage{amsmath}

\usepackage{color}
\definecolor{blue}{rgb}{0,0,1}
\definecolor{grey}{rgb}{0.6,0.6,0.6}

\def \ha{\hat{a}}
\def \hadag{\hat{a}^\dagger}

\def \hrho{\hat{\rho}}
\def \hd{\hat{d}}
\def \hddag{\hat{d}^\dagger}

\def \trho{\hat{\tilde{\rho}}_{\uparrow \downarrow}}

\def \ua{\uparrow}
\def \da{\downarrow}
\def \hH{\hat{H}}

\begin{document}

%\preprint{}

%\title{Qubit detection with a Nonlinear Cavity Detector Beyond Linear Response}
\title{Weak Qubit Measurement with a Nonlinear Cavity: Beyond Perturbation Theory}


\author{C. Laflamme}
\affiliation{Physics Department, McGill University, Montreal,
Quebec, Canada H3A 2T8}
\author{A. A. Clerk}
\affiliation{Physics Department, McGill University, Montreal,
Quebec, Canada H3A 2T8}
%\thanks{}


\date{\today}

\begin{abstract}
We analyze the use of a driven nonlinear cavity to make a weak continuous measurement of 
a dispersively-coupled qubit. We calculate
the backaction dephasing rate and measurement rate 
beyond leading-order perturbation theory using a phase-space approach which accounts for 
cavity noise squeezing.     
Surprisingly, we find that increasing the coupling strength beyond the regime describable by leading-order perturbation theory
(i.e.~linear response) allows one to come
significantly closer to the quantum limit on the measurement efficiency.
We interpret this behaviour in terms of the non-Gaussian photon number fluctuations
of the nonlinear cavity. 
Our results are relevant to recent experiments using superconducting microwave circuits to 
study quantum measurement.  
\end{abstract}

%\keywords{}

\maketitle


{\it Introduction-}
There is considerable interest in exploiting continuous weak quantum measurements
for the detection of fundamental quantum behavior as well as for quantum information 
processing \cite{Devoret00, Makhlin01, Clerk2010}.
By weak measurement, we mean the generic situation where the signal produced in the detector by the 
measured system is small compared to intrinsic output noise, and thus information is obtained only gradually in time.
%A weak measurement results when the coupling strength between the system of interest and detector is too weak to make a projective, instantaneous measurement, and to gain appreciable information about the system one must integrating the output signal over time \cite{Clerk2010}. 
Such measurements are ultimately constrained by the Heisenberg uncertainty principle, which dictates that the backaction disturbance of the system by the detector cannot be arbitrarily small, but is instead bounded by the rate at which information is acquired \cite{Braginsky92, Devoret00, Korotkov01c, Makhlin01, Clerk2010}. 
% In the specific case of QND qubit detection which we will focus on, the measured quantity is the $\sigma_z$ operator of a qubit, and the backaction corresponds to measurement-induced qubit dephasing.
Detectors capable of yielding an optimally small ratio of backaction-to-information gain are known as quantum-limited.  They are both of fundamental interest, and are also necessary if one wishes to implement 
continuous quantum feedback algorithms \cite{Wiseman93,Wiseman01,Korotkov05} or certain quantum error correction schemes \cite{Ahn2002}. 

\begin{figure}[t]
\includegraphics[width=0.99 \columnwidth]{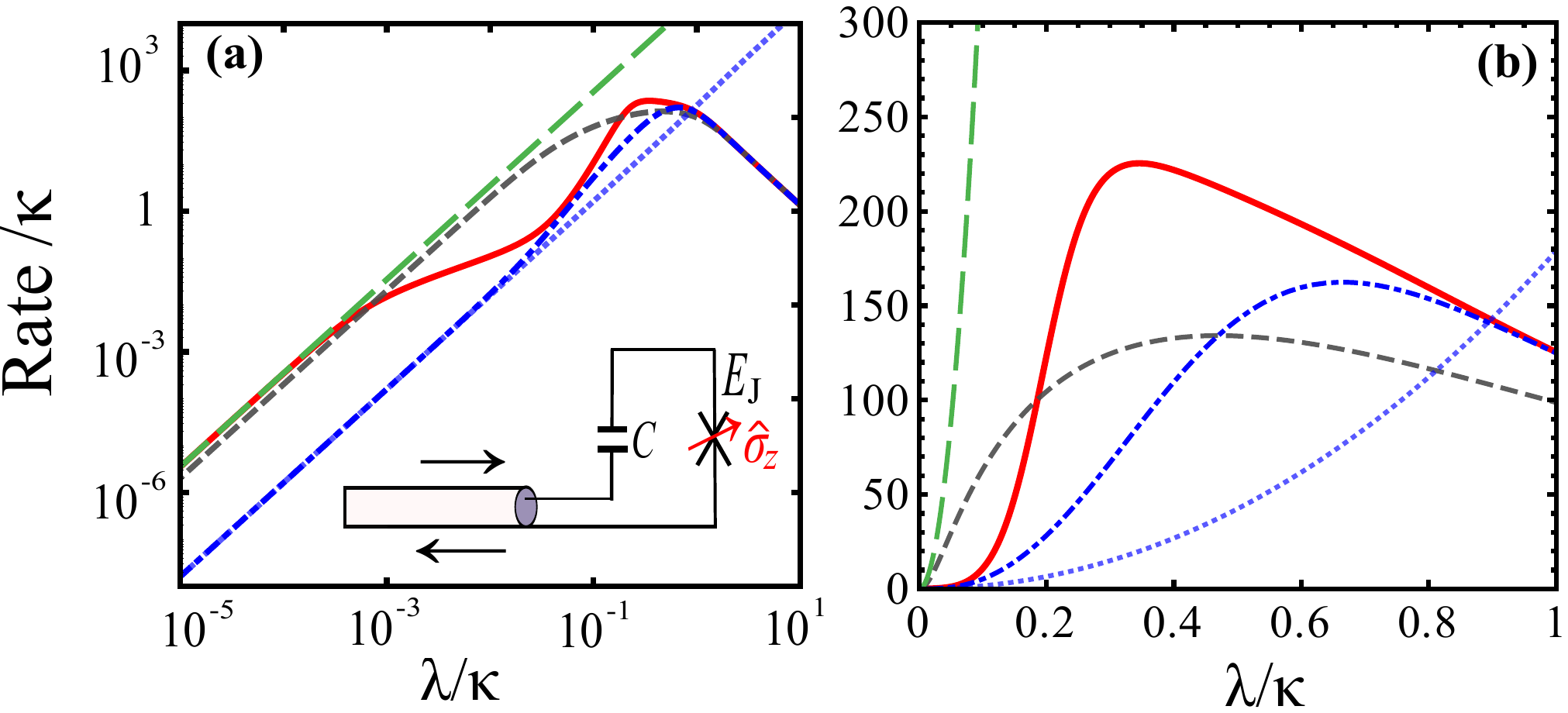}
\caption{(a) Inset: schematic showing a Josephson-junction circuit (i.e.~nonlinear cavity) dispersively coupled to a qubit.  
Main: Measurement rate and dephasing rate versus qubit coupling strength $\lambda$, in units of the cavity damping rate $\kappa$, using logarithmic axes.
The blue (dot-dashed) curve is the full measurement rate $\Gamma_{\rm meas}$, while the light-blue (doted) curve is the linear-response
approximation to $\Gamma_{\rm meas}$.
The red (solid) line is the dephasing rate 
$\Gamma_{\varphi}$ as obtained from the full
theory presented in the main text.  The remaining lines are $\Gamma_{\varphi}$ calculated within less rigorous approximations: the grey (dotted) 
curve is the 
linear-cavity formula $\Gamma_{\varphi,0}$ (c.f. Eq.~(\ref{eq:me_zerotemp_simple})) and green (dashed) curve is leading-order perturbation theory.
Parameters are
$\Lambda = 10^{-3}\kappa, f_0=0.75f_{\rm bif}, \bar{n}\sim 200, \Delta = \Delta_{\rm bif}$ where $f_{\rm bif}$ and 
$\Delta_{\rm bif}$ are the driving force amplitude and detuning at the cavity bifurcation.  The parametric photon number gain is $G \sim 10^2$.  One clearly sees that for moderate couplings, the dephasing rate 
is strongly suppressed compared to the perturbative result.
(b) Same, but using linear axes.}   
\label{fig:Rates}
\end{figure}

Not surprisingly, weak measurements are usually analyzed in the limit of a system-detector coupling small enough that 
leading-order perturbation theory in the coupling applies; in this standard regime, the quantum limit reduces to a constraint on the noise properties of the detector \cite{Braginsky92, Clerk2010}. 
Here, we focus on an alternate regime, where the detector-system coupling is still weak enough that information is obtained gradually in time, but not so weak that leading-order perturbation theory is sufficient.   
This regime of a ``weak-but-not-too-weak" measurement has recently been achieved in experiments using a driven, nonlinear superconducting microwave cavity to measure the state of a superconducting qubit 
\cite{Esteve10}.  As with experiments using linear microwave cavities \cite{Wallraff04,Blais04}, the qubit is dispersively coupled to the cavity, meaning that the cavity frequency depends on the qubit state; by monitoring the phase of reflected microwaves from the cavity, one can monitor the qubit state. Introducing a nonlinearity in the cavity via a Josephson junction
(see Fig.~\ref{fig:Rates}) allows one to operate the cavity detector close to a point of bifurcation, where the state of the driven cavity is an extremely sensitive (but still single-valued) function of its frequency.  
The enhanced sensitivity of this regime naturally leads to conditions where the measurement is weak, but the qubit-detector coupling cannot be treated perturbatively.  While information gain is enhanced here, the question remains whether this speedup comes at the cost of deviating from the quantum limit (i.e.~excess backaction dephasing).

In this Letter, we present an analytic theory describing weak measurement of a qubit with a nonlinear cavity operated close to a point of bifurcation.  Our non-perturbative approach accounts for the non-trivial cavity noise physics associated with the nonlinearity.  We find that 
the information-gain to state-disturbance ratio is a strong function of the qubit-detector coupling strength.  In the limit of an extremely weak qubit-detector coupling,  we recover previous perturbative results \cite{Wilhelm10,Laflamme2011,Bertet12,Boissonneault2012}, which indicate a large deviation from the quantum limit:  the backaction dephasing rate is a large factor $G$ greater than the
rate of information acquisition (the measurement rate), where $G \gg 1$ is the parametric photon-number gain associated with the driven nonlinear cavity.  Increasing the coupling beyond the perturbative regime, we find remarkably that the dephasing rate is greatly suppressed compared to the leading-order prediction.  This allows one to approach the quantum limit to within a factor of order unity. 
%\cut{This effective suppression of backaction with coupling strength can be naturally understood in terms of the non-Gaussian statistics of photon number fluctuations of the driven cavity.}  
 Our approach provides a general framework for investigating quantum measurement with driven nonlinear systems beyond weak coupling.
%\ACcomment{Hmm, we need to define measurement rate and dephasing rate somewhere before this paragraph.}

Note that the backaction of a nonlinear cavity detector was also considered by Boissonneault and co-workers \cite{Boissonneault2012,Bertet12, Esteve10}.  They described backaction dephasing beyond lowest-order in the coupling by approximating the
state of the driven cavity state {\it conditioned} on the qubit state to be a simple coherent state; this is only valid for operating points far from a bifurcation, where the detector has a relatively small gain.  We show that close to a bifurcation (where the cavity exhibits parametric gain and squeezing), this approach does not accurately capture the coupling dependence of the backaction dephasing.

{\it Model-- }
We consider a qubit coupled dispersively to a single-sided  nonlinear cavity. While our approach 
applies to an arbitrary nonlinearity, we focus here on the 
typical experimental situation \cite{Esteve10,Vijay10} where a Kerr-type nonlinearity dominates.  
Working in a frame rotating at the cavity drive frequency $\omega_{\rm d}$, the Hamiltonian is ($\hbar =1$)
\begin{equation}
\label{eq:Hamiltonian}
\hat{H}_{\rm sys}=-\Delta \hadag\ha -\Lambda\hadag\hadag\ha\ha +\hat{H}_{\kappa}+\hat{H}_{\rm qb}+\hat{H}_{\rm int},
\end{equation}
where $\Delta=\omega_{\rm d}-\omega_{\rm cav}$ is the detuning between the cavity drive and resonance frequency, $\Lambda$ is the Kerr constant, 
%$\ha/\hadag$ is the cavity lowering/raising operator, 
$\hat{H}_\kappa$ describes the cavity damping (with rate $\kappa$) and driving due to coupling to external modes,
and $\hat{H}_{\rm qb} = \Omega \hat{\sigma}_z$ is the qubit Hamiltonian. The dispersive QND qubit-cavity coupling Hamiltonian is $\hat{H}_{\rm int}=\lambda \hat{\sigma}_z\hadag \ha$, where the coupling strength $\lambda$ sets the qubit-dependent cavity frequency shift.  We will take the cavity to be at zero temperature, and ignore any intrinsic qubit dissipation, as we are interested only in the measurement backaction. 

As discussed, making a weak $\sigma_z$ measurement of the qubit involves strongly driving the cavity while monitoring the reflected light from the cavity via a homodyne measurement; the two possible values of $\sigma_z$ will lead to two different average homodyne currents, which as time progresses can be resolved above the intrinsic noise in these currents \cite{Blais04,Clerk2010}.  We will focus exclusively on a weak nonlinearity $\Lambda \ll \kappa$ and a strong drive amplitude, such that the stationary average cavity photon number 
$\langle \hat{a}^\dagger \hat{a} \rangle \gg 1$ regardless of the initial qubit state.  Apart from this constraint, we will not place any other restrictions on how small the qubit-cavity coupling $\lambda$ must be.

\begin{figure}[t]
\includegraphics[width=0.99 \columnwidth]{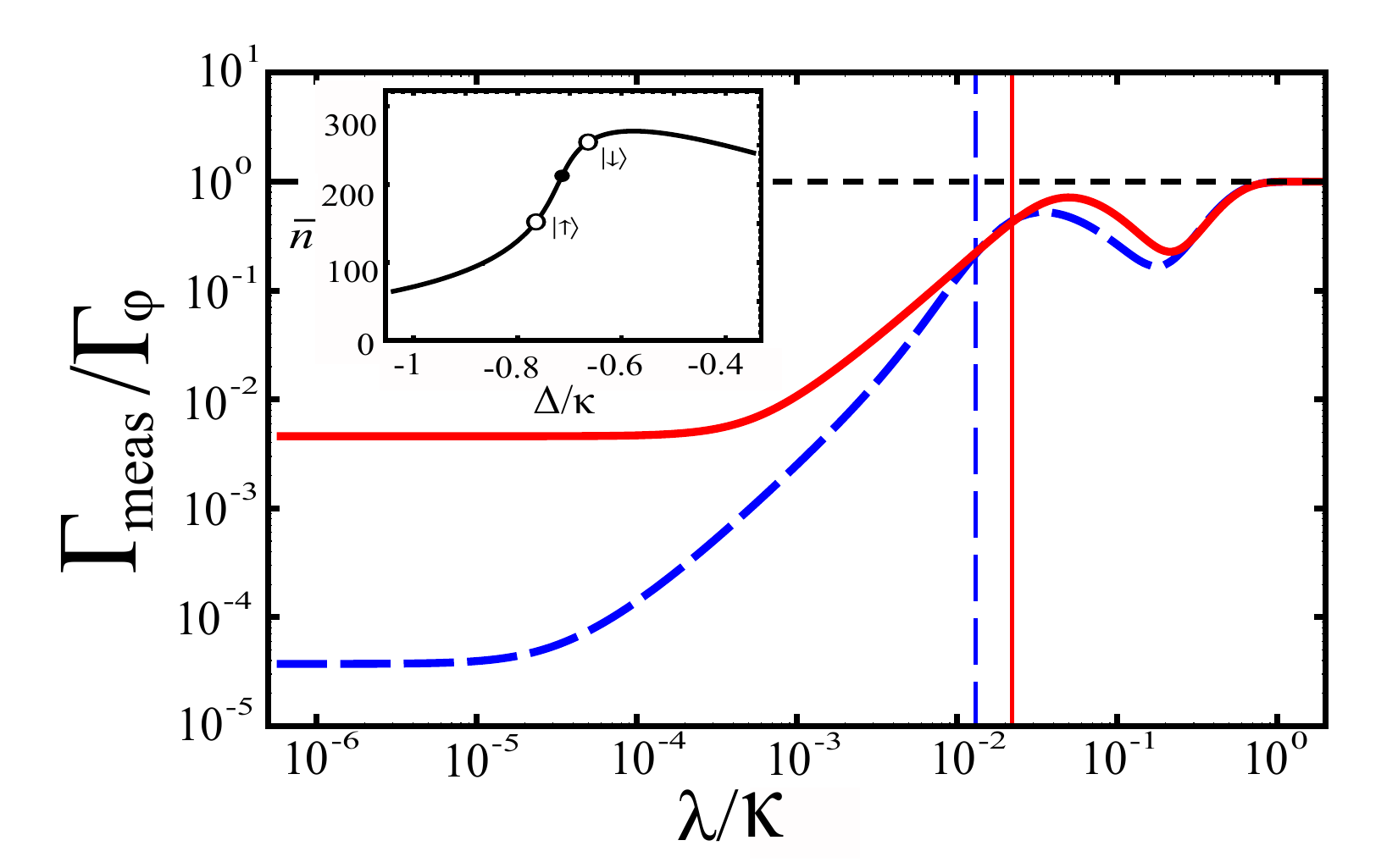}
\caption{Measurement efficiency ratio $\chi \equiv \Gamma_{\rm meas} / \Gamma_{\varphi}$ as a function of coupling strength for the same parameters
as Fig.~1 ($G=10^2$, red-solid curve), and for an operating point closer to the bifurcation yielding $G \sim 10^3$ (blue dashed solid curve); the
quantum limit is $\chi = 1$ (dashed-black line).  Vertical lines indicate 
the maximum $\lambda$ for which the measurement time $1 / \Gamma_{\rm meas}$ is longer than the detector response time 
$1 / \kappa_{\rm slow} \sim \sqrt{G} / \kappa$.
For $\lambda \rightarrow 0$, one misses the quantum limit by a large amount: $\chi \sim 1/G$.  However,
a small increase in coupling greatly improves this efficiency ratio.  
Inset: Average cavity photon number versus drive detuning $\Delta$ (parameters as in Fig. 1).  The black point indicates the chosen working point.
The two qubit states lead to two different effective values of $\Delta$; these are shown as white circles for $\lambda = 0.05 \kappa$.  
}
\label{fig:QLPlot}
\end{figure}

{\it Dephasing rate--}
We first calculate the backaction dephasing of the qubit that occurs during such a measurement; this dephasing is a direct consequence of the intracavity photon-number fluctuations.  If the qubit is initially in a $\sigma_z$ eigenstate, it will remain in this state (due to the QND nature of the measurement); thus, from the cavity's perspective, the two qubit eigenstates simply correspond to a shift of the cavity resonance frequency by either $\pm \lambda$. 
In each case, the classically-expected cavity amplitude $\alpha_\sigma$ ($\sigma = \ua, \da$) will be given by
\begin{eqnarray}
\label{eq:classical_value}
\left[-\frac{\kappa}{2}+i(\Delta \mp \lambda+2\Lambda|\alpha_{\ua/\da}|^2)\right]\alpha_{\ua / \da}&=&-if_0,
\end{eqnarray} 
where 
$f_0$ is the amplitude of the cavity drive.  Note that we focus on driving strengths small enough that we are below the bifurcation, i.e. there is only one classical solution $\alpha$ for a given $\Delta$.  By now writing $\hat{a} = \hat{d} + \alpha_\sigma$ and using the fact that $|\alpha_\sigma| \gg 1$, we can approximate the cavity Hamiltonian corresponding to each qubit eigenstate by only keeping terms that are at most quadratic in $\hat{d}, \hat{d}^\dag$.
We thus obtain {\it two} linearized cavity Hamiltonians, corresponding to the two qubit states.  Each has the general form of a 
degenerate parametric amplifier (DPA) driven by an off-resonant pump \cite{Carmichael84, Yurke06, Laflamme2011}:
\begin{equation}
\label{eq:two_Hamiltonians}
\hat{H}_{\sigma}=-\tilde{\Delta}_{\sigma} \hddag\hd +\frac{i}{2}(\tilde{g}_{\sigma}\hddag\hddag-\tilde{g}_{\sigma}^*\hd\hd),
\end{equation}
where $\tilde{\Delta}_{\ua / \da}= \Delta \mp \lambda + 4 |\alpha_{\ua / \da}|^2 \Lambda$ is the effective pump detuning, and $\tilde{g}_{\sigma} = -2 i \alpha_{\sigma}^2 \Lambda$ is the parametric strength.  As one approaches a point of bifurcation in the cavity (e.g.~by increasing the drive strength $f_0$), the corresponding DPA Hamiltonian approaches the threshold of self-oscillation \cite{Laflamme2011} 
(see inset of Fig.~\ref{fig:QLPlot}).  For such operating points, incident waves on the cavity in the appropriate quadrature will be strongly amplified; this amplification is described by a photon number gain $G$ which diverges
as one approaches the bifurcation, as well as a narrow-bandwidth $\kappa_{\rm slow} \sim \kappa / \sqrt{G}$
\cite{Laflamme2011,Vijay10,Yurke06}.

While the cavity evolution is easy to understand when the qubit is initially in a $\sigma_z$ eigenstate, to calculate the dephasing rate we need to understand the cavity dynamics when the qubit is in a superposition of its eigenstates.  We focus on the long-time qubit dephasing rate, which is defined as usual in terms of the decay of the qubit's off-diagonal density matrix elements in the long-time limit:  $ -| \ln {\rm Tr} \left( \hrho | \da \rangle \langle \ua | \right) | / t \rightarrow \Gamma_{\varphi}$, 
where $\hrho$ is the density matrix describing the full system.

To proceed, we first introduce  $\hrho_{\ua \da} =  {\rm Tr_{qb,bath}}\big(\hat{\rho}|\downarrow\rangle \langle \uparrow |\big)$, where the trace is over the qubit and cavity bath degrees of freedom.  This is an operator acting in the cavity Hilbert space; its trace yields the off-diagonal element of the qubit density matrix, and hence
can be used to obtain $\Gamma_{\varphi}$.  We further transform $\hrho_{\ua \da}$ by displacing away the two 
stationary classical cavity amplitudes $\alpha_\sigma$ associated with each qubit state.
%, as well as scale out the dephasing expected from a {\it linear} cavity, $\Gamma_{\varphi,0} = (\kappa/2) |\alpha_\ua - \alpha_\da|^2$ [CITE].  
We thus obtain a 
operator $\trho$:
\begin{equation}
%	\trho e^{-\Gamma_{\varphi,0}t} \equiv \hat{D}(\alpha_\uparrow)\hat{\rho}_{\uparrow \downarrow}\hat{D}^\dagger(\alpha_\downarrow),
	\trho(t)  \equiv \hat{D}(-\alpha_\uparrow)\hat{\rho}_{\uparrow \downarrow}(t) \hat{D}^\dagger(-\alpha_\downarrow),
\end{equation}
%\begin{equation}
%	\trho e^{-\Gamma_{\varphi,\alpha}t} \equiv \hat{D}(\alpha_\uparrow) \cdot
%	 {\rm Tr_{qb}}\big(\hat{\rho}|\downarrow\rangle \langle \uparrow |\big) \cdot 
%	\hat{D}^\dagger(\alpha_\downarrow),
%\end{equation}
where
% $ \hat{D}(\alpha)\hat{a}\hat{D}^\dagger(\alpha) =\hat{a}+\alpha$ 
$ \hat{D}(\alpha)= \exp( \alpha \ha^\dag - h.c. )$ is the cavity displacement operator.  One can show that in the long-time limit, the exponential decay of $\rm{Tr} \, \trho(t)$ also yields the dephasing rate $\Gamma_{\varphi}$ \cite{EPAPS}.

It is now straightforward to rigorously derive the evolution equation of $\trho$, starting from the standard Linblad master equation describing the evolution of the cavity-plus-qubit density matrix \cite{EPAPS} (see also \cite{Boissonneault2012}):  
\begin{eqnarray}
\label{eq:me_zerotemp_simple}
%\frac{\partial }{\partial t}\trho &=&-i(\hat{H}_{\uparrow}\trho -\trho\hat{H}_{\downarrow})+\kappa\mathcal{D}[\hat{d}]\trho -\frac{\kappa}{2}|\alpha_\uparrow - \alpha_\downarrow|^2\trho\nonumber\\ 
%&&+\kappa(\alpha_\uparrow-\alpha_\downarrow)\trho\hddag -\kappa(\alpha^*_\uparrow-\alpha^*_\downarrow)\hd\trho 
	\frac{\partial }{\partial t}\trho 
		&=&	 \kappa\mathcal{D}[\hat{d}]\trho 
				-i(\hat{H}_{\uparrow}\trho -\trho\hat{H}_{\downarrow})
				- \Gamma_{\varphi,0} \trho 
	\nonumber 
	\\
	&&	+ \kappa
		\left[ 
			(\alpha_\uparrow-\alpha_\downarrow)\trho\hddag - (\alpha^*_\uparrow-\alpha^*_\downarrow)\hd \trho
		\right].
%		\nonumber
\end{eqnarray}
Here $\mathcal{D}[\hat{d}]\trho =\hd\trho\hddag - (\hddag\hd\trho + \trho \hddag\hd)/2$ is the standard Lindblad super-operator
describing cavity damping.  The second and third terms in Eq.~(\ref{eq:me_zerotemp_simple}) correspond to the Hamiltonian evolution of the cavity in our doubly-displaced frame, where we have used $|\alpha_\sigma| \gg 1$ to linearize the   
two cavity Hamiltonians $\hH_{\sigma}$ (c.f.~Eq.~(\ref{eq:two_Hamiltonians})).  
The remaining terms on the RHS of Eq.~(\ref{eq:me_zerotemp_simple}) describe decoherence of the qubit 
resulting from the combination of the cavity drive and cavity dissipation, with $\Gamma_{\varphi,0} = (\kappa/2) |\alpha_\ua - \alpha_\da|^2$.  

 For a linear cavity, the terms on the last line of Eq.~(\ref{eq:me_zerotemp_simple}) play no role, and the backaction dephasing rate is given completely by $\Gamma_{\varphi,0}$ (i.e.~by the distinguishability of the two classical cavity amplitudes) \cite{Gambetta06}.
%%$\Gamma_{\varphi,0} = (\kappa/2) |\alpha_\ua - \alpha_\da|^2$ is an effective decay rate arising from the difference of the two classical cavity amplitudes; in the case of a purely linear cavity, this expression yields the exact backaction dephasing rate at zero temperature \cite{Gambetta06}.    
%%Finally, the last line of Eq.~(\ref{eq:me_zerotemp_simple}) corresponds to a modification of the effective dephasing rate $\Gamma_{\varphi}$ 
% to the tendency of the cavity dissipation to correlate the time evolution of the left and right of the density matrix, something that is transparent in the so-called Keldysh technique [CITE].  
%Note that the the fact that Eq.~(\ref{eq:me_zerotemp_simple}) is not of Linblad form is of no concern, as $\trho$ is not required to be a positive-definite operator.
For our case of a nonlinear cavity, the same is true {\it if} one neglects the parametric amplification terms in $\hH_{\rm \sigma}$ (proptional to $\hd^2$ and $(\hd^\dag)^2$), as then 
$\trho(t) = C \exp(- \Gamma_{\varphi,0} t) |0 \rangle \langle 0 |$ (where $|0 \rangle$ is the vacuum state and $C$ a constant) trivially solves Eq.~(\ref{eq:me_zerotemp_simple}).  This is equivalent to finding that 
(in the long time limit) 
the cavity state conditioned on the qubit is a coherent state $| \alpha_\sigma \rangle$.
In this approximation, the backaction dephasing rate is given
completely by the linear-cavity formula $\Gamma_{\varphi,0}$ \cite{Boissonneault2012,Bertet12, Esteve10}.  However, such an approximation completely neglects the squeezing of noise by the nonlinear cavity.  It is thus only valid for cavity parameters that are extremely far from any bifurcation, in regimes where the nonlinear cavity closely resembles a linear cavity.
 
%We stress that this linearization procedure {\it does not} require $|\alpha_\ua - \alpha_\da|$ to be small, as we are effective linearizing the dynamics separately about each of the two conditional cavity amplitudes.  
%\ACcomment{Could save space and just introduce linearization procedure once, say here.}
%%Rather, we only require the weaker condition 
%%$|\alpha_\sigma| \gg 1$.

Given the importance of noise squeezing, we go beyond the above approximation by retaining all terms in Eq.~(\ref{eq:me_zerotemp_simple}).  Defining $\nu(t) =  -\ln {\rm Tr} \trho(t)$, the long-time backaction dephasing rate will be given by 
$\Gamma_{\varphi} = \textrm{lim }_{t \rightarrow \infty}  \textrm{Re } \nu(t) / t$.  
%Taking the trace of Eq.~(\ref{eq:me_zerotemp_simple}) and defining $\delta \alpha  = \alpha_\ua - \alpha_\da$ yields
Setting $\delta \alpha  = \alpha_\ua - \alpha_\da$, the trace of Eq.~(\ref{eq:me_zerotemp_simple}) yields
\begin{eqnarray}
	\dot{\nu} & = & \Gamma_{\varphi,0} 
		-i  \left \langle \hH_{\ua} - \hH_{\da} \right \rangle_{\ua \da} 
	+ \kappa 
		\left \langle \delta \alpha \cdot \hd^\dagger - h.c. \right \rangle_{\ua \da},
	\label{eq:FullDephasing}
\end{eqnarray}
where we have defined the quasi-expectation value  $\langle \hat{\mathcal{O}}\rangle_{\ua \da} = {\rm Tr}( \hat{\mathcal{O}}\trho) / {\rm Tr}( \trho)$.   As $\trho$ is not a true density matrix, the quasi-expectation of a Hermitian operator can be complex, and hence the second third terms above can contribute to the backaction dephasing.  

We now use the fact that Eq.~(\ref{eq:me_zerotemp_simple}) only involves terms that are at most quadratic in $\hd$,$\hd^\dag$, and hence can be solved exactly by a $\trho$ which has a a Gaussian form (i.e.~its phase space representation is Gaussian \cite{EPAPS}).  
% the initial cavity state (before the drive is turned on) is a Gaussian state (e.g.~the ground state with no photons), implying that $\trho(t=0)$ also has a Gaussian form.  Further, Eq.~(\ref{eq:me_zerotemp_simple}) will keep $\trho$ Gaussian at all later times.  
Eq.~(\ref{eq:me_zerotemp_simple}) thus reduces to a closed set of evolution equations for the quasi-means and covariances of $\hd, \hd^\dagger$ (see \cite{EPAPS} for details).  Solving these and substituting into Eq.~(\ref{eq:FullDephasing}) directly gives $\dot{\nu}$ and thus the dephasing rate.  We stress that this approach is not perturbative in the coupling $\lambda$, and it does not neglect the noise squeezing expected near a cavity bifurcation.  A similar procedure can be used to calculate the backaction dephasing of a linear cavity subject to both quantum and thermal noise \cite{Clerk07}.  

%\begin{eqnarray}
%\label{eq:EOM_nu}
%	\dot{\nu}	&=&	
%		-i\mathrm{Re}[\tilde{g}_\uparrow -\tilde{g}_\downarrow] (\bar{x}\bar{p}+\sigma_{xp})
%	\nonumber \\
%	&&
%		+i\frac{\mathrm{Im}[\tilde{g}_\uparrow 
%		-\tilde{g}_\downarrow]}{2}(\bar{p}^2-\bar{x}^2+\sigma_p-\sigma_x)
%		\nonumber	\\ 
%	&&
%		-i\frac{(\tdelup-\tdeld)}{2}(\bar{x}^2+\bar{p}^2+\sigma_x+\sigma_p-1)
%	\nonumber	\\
%	&&
%		+\kappa(\alpha_\uparrow-\alpha_\downarrow)\frac{(-\bar{x}+i\bar{p})}{\sqrt{2}}+
%		\kappa(\alpha^*_\uparrow-\alpha^*_\downarrow)\frac{(\bar{x}+i\bar{p})}{\sqrt{2}}, 
%	\nonumber	\\
%\end{eqnarray}
%\ACcomment{Need to check signs etc. in this equation!!!!}
%One can rigorously show that the Heisenberg limit
%\begin{equation}
%\label{eq:QL_def}
%\Gamma_\varphi \geq \Gamma_{\rm meas}
%\end{equation} 
%(where $\Gamma_{\rm meas}$ is defined as per Eq.~(\ref{eq:meas_rate_def})) continues to hold beyond the limit of a vanishingly weak qubit-detector coupling $\lambda$.

To gain insight on the effect of increasing $\lambda$, we first use the above approach to calculate $\Gamma_{\varphi}$ to order $\lambda^4$.  For a cavity detuning $\Delta$ and drive $f_0$ chosen to be close to the bifurcation point, one finds
%\begin{equation}
%\label{eq:dephasing_series}
%	\Gamma_{\varphi}  =    
%		\frac{\lambda^2}{\kappa} \left(
%			\frac{2}{3} |\alpha_0|^2 G + \frac{1}{9} G^{5/2} \right) 
%			-  \frac{ \lambda^4}{\kappa^3} 
%			\left( \frac{32}{27}  |\alpha_0|^2 G^4 
%			+ \frac{5}{81} G^{11/2} \right)
%			 + \mathcal{O}(\lambda^6),
%\end{equation}
\begin{equation}
	\label{eq:dephasing_series}
	\Gamma_{\varphi}   \simeq    
		 \frac{\lambda^2 \bar{n} }{\kappa} G 
		\left[ 
			\left(
				\frac{2}{3} +
			 	 \frac{ G^{3/2}}{9 \bar{n}} 
			\right) 
			-  \frac{ \lambda^2}{\kappa^2}
			\left( 
				 \frac{32}{27}   G^3 
					+ \frac{5}{81} 
					 \frac{ G^{9/2}}{\bar{n}}  
			\right)
		\right],
%			 + \mathcal{O}(\lambda^6),
\end{equation}
where $\bar{n} = |\alpha_0|^2$, 
$\alpha_0$ is the zero-coupling classical cavity amplitude (i.e. solution to Eq.~(\ref{eq:classical_value}) at $\lambda=0$),  
and $G \gg 1$ is the parametric
photon-number gain (see Ref.~\cite{Laflamme2011}).
At each order in $\lambda$, we have retained the leading terms in $\bar{n}$ and $G$.
The first term here ($\propto \lambda^2 \bar{n}$)
reproduces the results of Ref.~\cite{Esteve10,Laflamme2011,Boissonneault2012} 
and arises solely from the linear-cavity dephasing rate $\Gamma_{\varphi,0}$ in Eq.~(\ref{eq:FullDephasing}).  The second term (also order $\lambda^2$) is missed if one linearizes
the qubit-cavity interaction, or makes the approximation 
$\Gamma_{\varphi} = \Gamma_{\varphi,0}$.  More interesting are the leading $\lambda^4$ terms.  Surprisingly, this correction is {\it negative}, suggesting the possibility of a relative suppression of dephasing with increased coupling (relative to the lowest-order-in-$\lambda$ expression).  The leading $\lambda^4$ corrections are completely due to the last line of Eq.~(\ref{eq:FullDephasing}), terms that would vanish if one ignored the squeezing of noise near the bifurcation.  As such, the approximation $\Gamma_{\varphi} \simeq\Gamma_{\varphi,0}$ would predict both an incorrect sign and scaling with $G$ of this term.
 
To see the full consequence of this behaviour,  we numerically solve Eq.~(\ref{eq:FullDephasing}) and the ODE's determining the needed averages; we use parameters corresponding to a weakly nonlinear cavity operated near a bifurcation point, 
similar to those realizable in experiment \cite{Esteve10,Vijay10}, and
well within the regime of validity of our theory.  In Fig.~\ref{fig:Rates}, one sees clearly that the backaction dephasing rate as a function of coupling (red) drops markedly below both the expectations from lowest-order perturbation theory (green), and below the linear-cavity formula $\Gamma_{\varphi,0}$ (grey).  We have confirmed that this behaviour is generic whenever one is close to a bifurcation in the cavity:  {\it higher-order-in-$\lambda$ terms yield a marked suppression of the dephasing rate.}
% compared both against the linear-cavity expression, and lowest-order perturbation theory.}
%
%In Fig.~(\ref{fig:displaced_dephasing}) we show the `complete' numerical solution to Eq.~(\ref{eq:EOM_nu}) using parameters chosen to be well within our approximation of $|\alpha_{\uparrow/\downarrow}| \gg 1$. The general behavior, however, is generic to within our approximation. 

%As expected by the perturbation expansion we find that the resulting dephasing rate is much lower than would be expected by extrapolating the linear response results. In addition to the suppression of the `complete' dephasing rate when compared to linear response, there is also a drastic difference when compared to the dephasing rate of two coherent pointer states (coherent state approximation). From these effects it is clear that including the effects of squeezed pointer states acts to reduce the backaction dephasing of the system. 

%%------------------------------------------------------------------------------------------------------------------------------------
%%------------------------------------------------------------------------------------------------------------------------------------
%%------------------------------------------------------------------------------------------------------------------------------------
%%------------------------------------------------------------------------------------------------------------------------------------

%\begin{figure}
%\includegraphics[angle=270,width=0.8 \columnwidth]{dephasing_plot}
%\caption{Dephasing rate for the complete analysis (red), coherent state approximation (black, dashed) and for linear response (blue, dot-dashed). Parameters: $\Lambda = 10^{-4}\kappa, f_0=0.95f_{\rm bif}, \bar{n}\sim 3 \times 10^3.$ The parametric gain is $G\sim3$.\label{fig:displaced_dephasing}.\\ {\bf Insert:} The average photon number $\bar{n}=|\alpha|^2$ for both qubit up (red,dot-dashed) as defined by Eq.~\ref{eq:classical_value} and for the qubit down (blue).}
%\end{figure}

%%------------------------------------------------------------------------------------------------------------------------------------
%%------------------------------------------------------------------------------------------------------------------------------------
%{\it Origin of suppression--}
For a heuristic understanding of the above behaviour, we
return to the physical picture that dephasing of the qubit is due to the photon-number fluctuations of the driven cavity.  Treating these fluctuations classically and defining $m(t) \equiv \int_0^t \mathrm{d} t' \,n(t')$ (where $n$ is the cavity photon number), one finds that the off-diagonal qubit density matrix is directly proportional to the characteristic function of the probability distribution of $m$ \cite{Levitov96,Clerk2011}.  As such, 
the long-time qubit dephasing rate can be expressed in terms of the even cumulants of $m$, $\langle\langle m^{2 j}\rangle\rangle$:
%\begin{equation}
%\label{eq:cumulant_expansion}
%\Gamma_\varphi  =  
%	\lim_{t \rightarrow \infty} \frac{1}{t} \left( 
%		2 \lambda^2\langle\langle \hat{m}^2\rangle\rangle - \frac{2}{3}\lambda^4\langle\langle \hat{m}^4\rangle\rangle +... \right),
%\end{equation}
\begin{equation}
\label{eq:cumulant_expansion}
\Gamma_\varphi  =  
	\lim_{t \rightarrow \infty} \frac{1}{t} 
		\sum_{j=1}^{\infty} (-1)^{j-1} \frac{ (2 \lambda)^{2j}}{( 2j )!} \langle \langle m^{2j} \rangle \rangle.
\end{equation}
This expansion also holds in the quantum case where $\hat{m}$ is an operator, if one now interprets the cumulants above using the standard Keldysh operator ordering \cite{Levitov96,Clerk2011}.  

Eq.~(\ref{eq:cumulant_expansion}) implies that terms of order $\lambda^4$ and higher in $\Gamma_{\varphi}$ are directly due to the non-Gaussian nature of intracavity photon number fluctuations.  In particular, having a negative contribution to $\Gamma_\varphi$
 at order $\lambda^4$ requires a positive kurtosis $\langle \langle m^4 \rangle \rangle$.  We note that even a driven linear cavity
in the classical limit has non-Gaussian intra-cavity photon number fluctuations and a positive kurtosis; this is a simple consequences of $n$ being the square of a Gaussian random variable (i.e.~the cavity amplitude) \cite{Clerk2011}.
A positive kurtosis indicates a distribution which is more peaked than a Gaussian, and hence noise that would generate
less dephasing than truly Gaussian noise.  In our nonlinear resonator, the non-Gaussian nature of the photon number fluctuations is strongly enhanced near bifurcation by the intrinsic nonlinearity of the system.  The kurtosis remains positive (like a linear resonator), but is much larger than would be expected for even a degenerate parametric amplifier near threshold \cite{EPAPS}.

{\it Measurement rate and quantum limit-- }
%Eq.~(\ref{eq:FullDephasing}) suggests that higher-order corrections to the dephasing rate will be important when $(\lambda / \kappa)^2 \sim 1/ G^2$.  For $G$ sufficiently large, there is a large range of coupling strengths where this condition is true, while at the same time, the measurement is weak, meaning that information is acquired slowly compared to internal detector timescales.
%%; this only requires 
%%$(\lambda / \kappa)^2 < 1 / (\sqrt{G} | \alpha_0 |^2)$.  
%
%
For a measurement that occurs slowly on detector timescales, we can characterize the information gain of the measurement by a single measurement rate
%Having found the backaction dephasing rate for arbitrary $\lambda$, we now wish to compare it against the 
%rate of information gain, the measurement rate.  
%Because of the intrinsic shot noise fluctuations in the homodyne current, it will take time to 
%distinguish the qubit states. 
$\Gamma_{\rm meas}$: how quickly
do the distributions of the output homodyne current corresponding to each qubit eigenstate ($\ua$ or $\da$) become distinguishable. 
% In our case,
%each of these distributions 
%
%
% measurement time $\tau_{\rm meas}$, namely how long must one wait before one can resolve  the two qubit energy eigenstates $| \uparrow \rangle$, $| \downarrow \rangle$ with a signal-to-noise ratio of 1.  
Generalizing the standard weak-coupling expression \cite{Devoret00,Makhlin01,Clerk2010} to a situation where the coupling is not perturbative but the measurement is still slow compared to internal detector timescales, we find (see \cite{EPAPS} for details)
\begin{equation}
\label{eq:meas_rate_def}
	\Gamma_{\rm meas} 
		= \frac{(\bar{I}_\uparrow - \bar{I}_\downarrow)^2}{4(S_{II,\uparrow}+S_{II,\downarrow})}.
\end{equation}
Here $\bar{I}_{\sigma}$ is the average stationary homodyne current when the qubit is in the state $\sigma = \uparrow, \downarrow$, and similarly, 
$S_{II,\sigma}$ is the zero-frequency spectral density of homodyne current fluctuations when the qubit is frozen in the 
state $\sigma$.  
%\textcolor{red}{In defining the measurement-rate, we have assumed that the measurement occurs slowly on the internal timescales of the detector, implying here that we need $\Gamma_{\rm meas} \ll \kappa / \sqrt{G} $.}
%Note that both the numerator and denominator of Eq.~(\ref{eq:meas_rate_def}) are functions of $\lambda$; in the $\lambda \rightarrow 0$ limit, $\Gamma_{\rm meas} \sim \lambda^2$.  
One can rigorously show that for arbitrary $\lambda$, the measurement efficiency ratio $\chi = \Gamma_{\rm meas} / \Gamma_{\varphi} \leq 1$ \cite{EPAPS}.

Using the linearized Hamiltonians $\hH_{\sigma}$ given in Eq.~(\ref{eq:two_Hamiltonians}) along with standard input-output theory \cite{WallsMilburn08} lets us evaluate Eq.~(\ref{eq:meas_rate_def}) for an arbitrary value of the coupling; comparing against the dephasing rate then allows us to investigate the behaviour of $\chi$ as a function of coupling.  One finds that similar to the linear-cavity dephasing rate $\Gamma_{\varphi,0}$, the measurement rate is largely determined by the classical amplitudes 
$\alpha_\sigma$, and is hence a far weaker function of $\lambda$ than the dephasing rate.  The result is that there is a range of $\lambda$ where higher-order
terms significantly suppress the dephasing rate (over the perturbative expression), whereas the measurement rate is still determined by the leading-order expression
(see Fig.~\ref{fig:Rates}a).  

%the measurement rate is a far weaker function of coupling than the dephasing rate.  
%Expanding in $\lambda$ and working near a bifurcation, one finds:
%\begin{equation}
%	\Gamma_{\rm meas} =  
%	\frac{  \lambda^2   \bar{n} }{2 \kappa} \left(
%		1 +    \frac{4}{9} G^{3/2}  \frac{\lambda^2}{\kappa^2} \right) + \mathcal{O}(\lambda^6)
%\end{equation}
%Thus, for 
%$(\lambda / \kappa)^2 \sim 1/ G^3$, the coupling is still weak enough that the measurement is slow compared to detector timescales (i.e.~$\Gamma_{\rm meas} < \kappa_{\rm slow} \sim \kappa/\sqrt{G}$), while at the same time higher-order-in-$\lambda$ corrections to the dephasing rate are significant (c.f.~Eq.~(\ref{eq:FullDephasing})). 

%%------------------------------------------------------------------------------------------------------------------------------------
%%------------------------------------------------------------------------------------------------------------------------------------
%%------------------------------------------------------------------------------------------------------------------------------------
%%------------------------------------------------------------------------------------------------------------------------------------

Shown in Fig.~\ref{fig:QLPlot} is $\chi$ versus $\lambda$ for the same parameters as in Fig.~\ref{fig:Rates}.  In the limit of a vanishing coupling strength, one deviates strongly from the quantum limit \cite{Laflamme2011}.  However, the effective suppression of dephasing that occurs with a modest increase of coupling brings one within a factor of order unity of the ultimate quantum limit bound $\chi  = 1$.  We find that this behaviour is generic for cavity operating points near bifurcation:  increasing the coupling $\lambda$ beyond the validity of leading-order perturbation theory allows one to make a weak measurement with a much higher efficiency than in the extreme weak coupling limit.  On a physical level, this is a direct result of the non-Gaussian nature of photon number fluctuations in the driven cavity (namely, the large positive kurtosis).

{\it Conclusions-} We have described a general method to calculate the measurement and dephasing rate of a qubit coupled to a nonlinear resonator that is not perturbative
in the qubit-detector coupling and which accounts for cavity noise squeezing.  By increasing the coupling to a regime where higher-order corrections are relevant, one can come significantly closer to the fundamental quantum limit on weak continuous qubit measurement.  

This work was supported by NSERC and CIFAR.

\section{}
\subsection{}
\subsubsection{}

\end{document}

\title{Supplementary material for ``Weak Qubit Measurement with a Nonlinear Cavity: Beyond Perturbation Theory"}
\author{Catherine Laflamme}
\affiliation{Department of Physics, McGill University, 3600 rue University, Montreal, QC
Canada H3A 2T8}
\author{Aashish Clerk}
\affiliation{Department of Physics, McGill University, 3600 rue
University, Montreal, QC Canada H3A 2T8}

\maketitle

\section{Equations of motion for doubly-displaced cavity density matrix}

The standard Lindblad master equation describing the evolution of the qubit-plus-cavity density matrix 
$\hat{\rho}$ is:
\begin{equation}
\label{eq:full_ME}
\frac{d}{dt}\hat{\rho}  =  -i[-\Delta \hadag\ha -\Lambda \hadag\hadag\ha\ha,\hat{\rho}]
	+if_0[\ha+\hadag,\hat{\rho}]
	-i \lambda [ \hadag\ha \cdot \hat{\sigma}_z \, ,\hat{\rho}] 
	+  \kappa \mathcal{D}[\ha]\hat{\rho},
\end{equation}
%of the reduced density matrix ($\hat{\rho}_{\uparrow \downarrow} = \langle \uparrow |\hat{\rho}|\downarrow \rangle$) at zero temperature is
%\begin{equation}
%\label{eq:full_ME}
%\frac{d}{dt}\hat{\rho}_{\uparrow \downarrow} =-i[-\Delta \hadag\ha -\Lambda \hadag\hadag\ha\ha,\hat{\rho}_{\uparrow \downarrow}]+if_0[\ha+\hadag,\hat{\rho}_{\uparrow \downarrow}]+\kappa \mathcal{D}[\ha]\hat{\rho}_{\uparrow \downarrow}-i\lambda\{\hadag\ha,\hat{\rho}_{\uparrow \downarrow}\},
%\end{equation}
where 
$\mathcal{D}[\ha]\hat{\rho}=\hd\hat{\rho} \hddag-(\hddag\hd\hat{\rho} +
\hat{\rho}\hddag\hd)/2$ is the standard Lindblad super-operator describing cavity damping.  Multiplying both sides of this
equation by $| \downarrow \rangle \langle \uparrow |$ and then tracing over the qubit degrees of freedom yields the
equation of motion for $\hat{\rho}_{\uparrow \downarrow}$.
We are interested in the evolution of the displaced density operator $\trho$ which is obtained by displacing
$\hat{\rho}_{\uparrow \downarrow}$ by different amounts on the left and right, as per Eq.~(4).  We thus displace Eq.~(\ref{eq:full_ME}) in the same way; this directly yields the equation of motion for $\trho$ given in Eq.~(5).  Note that while the underlying master equation in Eq.~(\ref{eq:full_ME}) is in Linblad form, the final evolution equation for the double-displaced density matrix $\trho$ is {\it not} in Linblad form (even though it followed directly from Eq.~(\ref{eq:full_ME})).  This is of no great concern, as $\trho$ is not required to be a positive operator.

%\ACcomment{You start with x and p being dimensionfull, and then suddenly get rid of the dimensions without explaining how.  Just set them
%to be dimensionless from the start.  I've tried to do this, but you'll need to check the normalization constants etc.}
To work with this displaced master equation we express $\trho$ in terms of its Wigner representation.  Defining the cavity quadratures
$\hat{x}= (\hd+\hddag) / \sqrt{2},\hat{p}=-i(\hd-\hddag) / \sqrt{2}$, the Wigner representation is defined by:
\begin{equation}
\label{eq:define_tildeW}
\widetilde{W}(x,p; t) = \frac{1}{ \pi }\int_{-\infty}^{\infty}  dy \, \langle x+y| \trho(t) | x-y\rangle e^{-2ipy } ,
\end{equation}
where $|x\rangle$ are eigenkets of $\hat{x}$.
For simplicity, we consider the Wigner function in Fourier space using the convention
\begin{eqnarray}
\widetilde{W}[k,q; t] &=& \int dx \, dp \, e^{i(kx+qp)} \widetilde{W}(x,p; t).
\end{eqnarray}
%\ACcomment{Are you sure that we should have $\sqrt{2 \pi}$'s here?  This isn't consistent with the next equation.}

As discussed in the main text, the equation of motion Eq.~(5) is solved exactly by taking $\trho$ to have a Gaussian form,  
i.e.~the Wigner representation $W[x,p]$ has a Gaussian form.  In Fourier space, we may thus write:
%
%
% at the initial time $\trho (t=0)$ will have a Gaussian form and since the master equation is quadratic in cavity operators, $\trho$ will remain in Gaussian form for later times . Beginning with $\widetilde{W}(x,p)$ as Gaussian we explicitly calculating the Fourier transform we obtain (now taking $x,p$ dimensionless)
\begin{eqnarray}
	\label{eq:tildeW_Fourier}
	\widetilde{W}[k,q; t] 
%	\frac{\widetilde{W}[k,q; t]}{\widetilde{W}[0,0;0]} 
		& = &
			e^{-\nu(t)}
			{\rm exp}\left[i(\bar{x}(t)k+\bar{p}(t)q)-\frac{1}{2}(k^2\sigma_x(t)+q^2\sigma_p(t)+2kq\sigma_{xp}(t))\right].
\end{eqnarray}
%Here, $\tilde{W}[0,0;0]$ simply characterizes the initial value of the qubit off-diagonal density matrix element.
Rewriting the evolution equation Eq.~(5) as a differential equation for $W(x,p)$ leads to a closed set of equations for the mean values $\bar{x}(t),\bar{p}(t)$,  the variances $\sigma_x(t),\sigma_p(t),\sigma_{xp}(t)$, and $\nu(t)$ (which sets the trace of $\trho$).  These means and covariances (as defined above) are equivalent to the definition in the main text involving a quasi-expectation value (see text after Eq.~(6)).
% For example, the mean value $\bar{x}$ is 
% \begin{equation}
%\bar{x}(t) ={\rm Tr}(\hx\trho)/{\rm Tr}(\trho).
%\end{equation} 

To present the equations of motion for the means and covariances which determine $\trho(t)$, it is useful to introduce parameters
characterizing the average and difference of the two cavity Hamiltonians  $\hat{H}_\uparrow, \hat{H}_\downarrow $ corresponding to the
two qubit eigenstates.  For each parameter $A$ in the cavity Hamiltonian of Eq.~(3) ($A = \tilde{g}, \tilde{\Delta}, \alpha$), we thus define:
\begin{eqnarray}
	\delta A &=& A_\uparrow - A_\downarrow \nonumber\\
	\bar{A}  &=& \frac{A_\uparrow + A_\downarrow}{2},
	\label{eq:BarDelta}
\end{eqnarray}
%\begin{eqnarray}
%	\delta \tilde{g} &=& \tilde{g}_\uparrow - \tilde{g}_\downarrow \nonumber\\
%	\bar{\tilde{g}}  &=& \tilde{g}_\uparrow + \tilde{g}_\downarrow,
%\end{eqnarray}
In the following equations, terms involving $\delta \tilde{g}$ and $\delta \tilde{\Delta}$ are a direct result of the different Hamiltonian evolution of the cavity associated with the two qubit states.  In contrast, terms involving $\delta \alpha$ arise from the non-Linblad dissipation terms in the last line of Eq.~(5); these terms make no contribution in the case of a linear cavity.

We first take the trace of Eq.~(5), which corresponds to setting $k=q=0$ in the Fourier-transformed phase space equation.  This then yields the equation of motion for the parameter $\nu(t)$ as given in Eq.~(6) in the main text.  Writing this explicitly, we have:
\begin{eqnarray}
\label{eq:EOM_nu}
\dot{\nu}  &=&  
	\Gamma_{\varphi,0} - i\frac{\delta\tilde{\Delta}}{2}(\bar{x}^2+\bar{p}^2+\sigma_x+\sigma_p-1) -i\sqrt{2}\kappa \mathrm{Im}[\delta\alpha]\bar{x} +i\sqrt{2}\kappa\mathrm{Re}[\delta\alpha]\bar{p} \nonumber \\&&\qquad\qquad \qquad-i\mathrm{Re}[\delta\tilde{g}] (\bar{x}\bar{p}+\sigma_{xp})+i\frac{\mathrm{Im}[\delta\tilde{\tilde{g}}]}{2}(\bar{p}^2-\bar{x}^2+\sigma_p-\sigma_x).
\end{eqnarray}
where $\Gamma_{\varphi,0} = \kappa | \delta \alpha |^2 / 2$.

Similarly, from Eq.~(5) we find that the equations for the mean values take the form:
\begin{eqnarray}
\label{eq:EOM_means}
\dot{\bar{x}}&=& \left(-\frac{\kappa}{2} + \mathrm{Re}[\bar{\tilde{g}}]\right)\bar{x} + \left(-\bar{\tilde{\Delta}} + \mathrm{Im}[\bar{\tilde{g}}]\right) \bar{p}+i\delta\tilde{\Delta}(\bar{x}\sigma_x+\bar{p}\sigma_{xp})+ i\frac{\kappa}{\sqrt{2}}  \mathrm{Im}[\delta \alpha] \left(2 \sigma_x-1\right) -i\sqrt{2}\kappa \mathrm{Re}[\delta \alpha]\sigma_{xp}\nonumber\\&&-i\mathrm{Re}[\delta \tilde{g}] (\bar{x}\sigma_{xp}+\bar{p}\sigma_x)+ i\mathrm{Im}[\delta \tilde{g}](\bar{x}\sigma_x-\bar{p}\sigma_{xp})\nonumber \\
%%%%%%%%%%%%
\dot{\bar{p}}&=& - \left(\frac{\kappa}{2}+ \mathrm{Re}[\bar{\tilde{g}}]\right)\bar{p}+\left(\bar{\tilde{\Delta}} + \mathrm{Im}[\bar{\tilde{g}}]\right)\bar{x}+ i\delta\tilde{\Delta}(\bar{p}\sigma_p+\bar{x}\sigma_{xp})- i\frac{\kappa}{\sqrt{2}}\mathrm{Re}[\delta \alpha]\left(2\sigma_p-1\right)+  i\sqrt{2}\kappa \mathrm{Im}[\delta \alpha]\sigma_{xp}\nonumber\\&&- i\mathrm{Im}[\delta \tilde{g}](\bar{p}\sigma_p-\bar{x}\sigma_{xp})-i\mathrm{Re}[\delta \tilde{g}] (\bar{p}\sigma_{xp}+\bar{x}\sigma_p).
\end{eqnarray}
In each of these equations, the first two terms are just the simple drift equations that one would obtain if the cavity was described by the average Hamiltonian $(\hat{H}_{\uparrow} + \hat{H}_{\downarrow}) / 2$.  The remaining terms result from the difference between these two conditional Hamiltonians, as well as from the non-Linblad dissipation terms in Eq.~(5).

Finally, from Eq.~(5), one finds that the equations of motion for the variances and covariances are
\begin{eqnarray}
\label{eq:EOM_variances}
%%%%%%%%%%%
\dot{\sigma}_x&=&\frac{\kappa}{2}+ (2\mathrm{Re}[\bar{\tilde{g}}]-\kappa -2i \mathrm{Re}[\delta \tilde{g}]\sigma_{xp} ) \sigma_x-(2\bar{\tilde{\Delta}}- 2\mathrm{Im}[\bar{\tilde{g}}])\sigma_{xp}+i \mathrm{Im}[\delta \tilde{g}](\sigma_x^2-\sigma_{xp}^2+1/4)+i\delta\tilde{\Delta}(\sigma_x^2+\sigma_{xp}^2-1/4) \nonumber\\
%%%%%%%%%%%%-
\dot{\sigma}_p&=&\frac{\kappa }{2}-(2\mathrm{Re}[\bar{\tilde{g}}]+\kappa-2i \mathrm{Re}[\delta \tilde{g}]\sigma_{xp})\sigma_p +(2\bar{\tilde{\Delta}}+ 2 \mathrm{Im}[\bar{\tilde{g}}])\sigma_{xp}-i\mathrm{Im}[\delta \tilde{g}](\sigma_{p}^2-\sigma_{xp}^2+1/4)+i\delta\tilde{\Delta}(\sigma_p^2+\sigma_{xp}^2-1/4)\nonumber\\ 
%%%%%%%%%%%%%
\dot{\sigma}_{xp}&=&\bar{\tilde{\Delta}}(\sigma_x-\sigma_p) + \mathrm{Im}[\bar{\tilde{g}}](\sigma_x+\sigma_p)-\kappa\sigma_{xp}\nonumber +i\delta\tilde{\Delta}(\sigma_p+\sigma_x)\sigma_{xp}-i\mathrm{Re}[\delta \tilde{g}] (\sigma_{p}\sigma_{x}+\sigma_{xp}^2+1/4)+i\mathrm{Im}[\delta \tilde{g}](\sigma_x-\sigma_p)\sigma_{xp}.\nonumber\\
\end{eqnarray}
Note these equations have the general form of a matrix Riccati equation.

\section{Obtaining the dephasing rate}

We focus on the long-time backaction dephasing rate of the qubit, which is defined as usual by the 
decay of the qubit off-diagonal density matrix:
\begin{equation}
	\Gamma_{\varphi} \equiv - \lim_{t \rightarrow \infty} 
		 \frac{  \ln \left| {\rm Tr} \left[ \hrho(t) | \da \rangle \langle \ua | \right] \right|  } { t  }.
	\label{eq:DephRateDefn}
\end{equation} 

As we now show, in the long-time limit, we can also obtain this dephasing rate by considering the decay of the doubly-displaced
density matrix $\trho$ defined in Eq.~(4):
\begin{equation}
\Gamma_{\varphi} 
	= - \lim_{t \rightarrow \infty} 
		 \frac{  \ln \left| {\rm Tr} \trho   \right|  } { t  }
	 = \lim_{t \rightarrow \infty} 
		\frac{ {\rm Re}[\nu] }{t}.
		\label{eq:UsefulDephRate} 
\end{equation}
To see this we consider the Wigner representation
$\tilde{W}(x,p)$ of $\trho$ (c.f.~Eq.~(\ref{eq:define_tildeW})), as well as that of $\hrho_{\uparrow\downarrow}$,
\begin{equation}
\label{eq:define_W}
W(x,p) = \frac{1}{ \pi}\int_{-\infty}^{\infty} \langle x+y| \hrho_{\uparrow\downarrow} | x-y\rangle e^{-2ipy } dy.
\end{equation}

Eq.~(4) of the main text tells us that the operators $\trho$ and $\hrho_{\uparrow\downarrow}$ are simply related by a pair of displacement transformations.  This immediately implies that their Wigner transforms are simply related by a displacement in phase space 
and an overall phase factor.
Writing $\alpha_\sigma = (u_\sigma + i v_\sigma) / \sqrt{2}$, and defining $\delta u, \delta v, \bar{u}, \bar{v}$ as per Eq.~(\ref{eq:BarDelta}), we have:
\begin{equation}
\label{eq:relate_wigner}
  \widetilde{W}(x,p) = e^{i\phi(x,p)} W( x+\bar{u},p+\bar{v}),
\end{equation}
where the (real) phase $\phi(x,p)$ is given by
\begin{equation}
  \phi(x,p)= (\delta v ( x+\bar{u})  + \delta u (p+\bar{v}))+(\bar{v} \delta u - \bar{u} \delta v).
\end{equation}
%For simplicity we work with the real and imaginary parts of the qubit-dependent classical cavity amplitudes 
%$\alpha_\sigma$ defined in Eq.~(2), 
%\ACcomment{Confusing:  earlier, the bar and delta notation is used with any factors of two, now you have factors of two.  Use the same notation in each case.  The difference shouldn't have a (1/2), the average should!}
%\begin{eqnarray}
%x_{0,\sigma}= \sqrt{2}{\rm Re}[\alpha_\sigma],\nonumber\\
%\bar{x}_{0}=\frac{1}{2}(x_{0,\uparrow}+x_{0,\downarrow}),\nonumber\\
%\delta x_{0}=\frac{1}{2}(x_{0,\uparrow}-x_{0,\downarrow}),
%\end{eqnarray}
%and analogous expressions for $\bar{p}_{0}$ and $\delta p_{0}$.
It thus immediately follows that since $\tilde{W}(x,p)$ has a Gaussian form, so will $W(x,p)$.  Using Eqs.~(\ref{eq:relate_wigner}) and 
(\ref{eq:tildeW_Fourier}), we thus find the Fourier 
transform of $W$ (defined as in Eq.~(\ref{eq:tildeW_Fourier})) is given by:
\begin{equation}
\label{eq:define_nuprime}
	W[k',q'] =
		e^{-\nu'}{\rm exp}
			\left[ 
				i((\bar{x}+\bar{u})k'+(\bar{p}+\bar{v})q')
				-\frac{1}{2}(k'^2\sigma_{ x}+q'^2\sigma_{p}+2k'q'\sigma_{xp})\right].
\end{equation} 
where the parameter $\nu'(t)$ is given by:
\begin{eqnarray}
\label{eq:relating_wigner_parameters}
	\nu'  & = &  
		\nu + 
		\left[\frac{ i}{2} ( \delta u^2  \sigma_p+ 
			\delta v^2 \sigma_x + 
			2 \delta v \cdot \delta u  \sigma_{xp}) - 
		\frac{1}{2}  
			\bar{u} \, \delta v  -\bar{x} \, \delta v  +\frac{1}{2} \bar{v} \, \delta u  -\bar{p} \, \delta u \right].
\label{eq:nuprime}
\end{eqnarray}

Note that the trace of $\rho_{\uparrow \downarrow}$ is just $W[0,0] = e^{-\nu'}$.
From the definition in Eq.~(\ref{eq:DephRateDefn}), the dephasing rate is thus rigorously given by:
\begin{equation}
\Gamma_{\varphi} 
	 = \lim_{t \rightarrow \infty} 
		\frac{ {\rm Re}[\nu']}{t}.
\end{equation}
While Eq.~(\ref{eq:nuprime}) implies that in general, $\nu(t) \neq \nu'(t)$ (i.e. $\trho$ and $\hrho_{\uparrow \downarrow}$ do not have the same trace), it also tells us that $\nu'(t) - \nu(t)$ tends to a constant in the long-time limit, as the covariances $\sigma_x, \sigma_p$ and $\sigma_{xp}$ all tend to a time-independent constant in this limit.  We have thus proved Eq.~(\ref{eq:UsefulDephRate}) which expresses the dephasing rate in terms of $\nu$ (i.e. the trace of $\trho$).  Thus, by solving Eqs.~(\ref{eq:EOM_nu}),(\ref{eq:EOM_means}) and (\ref{eq:EOM_variances}) we 
can directly calculate the backaction dephasing rate. 
%%%%%%%%%%%%%%%%%%%%%%%%

%%%%%%%%%%%%%%%%%%%%%%%%%%%%%%
%%%%%%%%%%%%%%%%%%%%%%%%%%%%%%
%%%%%%%%%%%%%%%%%%%%%%%%%%%%%%
\section{Dephasing rate and Kurtosis of intracavity photon number fluctuations}

As discussed in the main text, the strong suppression of backaction dephasing at order $\lambda^4$ can be directly attributed to a
strong, positive kurtosis (fourth cumulant) of the intracavity photon number fluctuations of the driven nonlinear cavity detector.  In this section we show that this enhanced kurtosis {\it cannot} be obtained if one approximates the nonlinear cavity by a detuned degenerate parametric amplifier.  Such an approximation is common, and is usually justified by first writing the cavity Hamiltonian in a frame where the classical
cavity amplitude $\alpha_0$ (given by Eq.~(2) with $\lambda=0$) is displaced to the origin.  Writing 
$\ha = \alpha_0 + \hd$, the full cavity Hamiltonian in this frame takes the form:
\begin{equation}
\label{eq:appendix_Hamiltonian}
	\hat{H} =-\tilde{\Delta}\hddag\hd +\frac{i}{2}(\tilde{g}\hddag\hddag-\tilde{g}^*\hd\hd)
	- \Lambda (2\alpha_0^* \hddag\hd\hd+2\alpha_0 \hddag\hddag\hd)-\Lambda\hddag\hddag\hd\hd.
\end{equation}
where $ \tilde{\Delta}=\Delta+4\Lambda|\alpha_0|^2$ is the effective drive detuning and 
$\tilde{g} = 2i\Lambda\alpha_0^2$ is the parametric interaction strength.
The standard approach is now to assume a large cavity photon number $\bar{n}=|\alpha_0|^2 $, 
and use this to drop terms that are
either cubic or quartic in the $\hd, \hd^\dag$ operators.  The resulting truncated Hamiltonian, $H_{\rm DPA}$, is that of a detuned, degenerate parametric amplifier (DPA) (i.e.~identical to Eq.~(3) of the main text with $\lambda = 0$).  We now show explicitly that this standard approach underestimates the magnitude the fourth cumulant of photon number fluctuations, even in the limit $\bar{n} \gg 1$.

The time-integral of the cavity photon number fluctuations is described by the operator
\begin{equation}
	\hat{m}  = \int_0^t d t' \left( \ha^\dag(t') \ha(t') - \bar{n} \right) =   
		\int_0^t d t' \;\left(\alpha_0 \hddag(t')+\alpha_0^*\hd(t') +\hddag(t') \hd(t')\right).
\end{equation}
Consider the fourth cumulant of $\hat{m}$, calculated using the approximate DPA Hamiltonian.  We would like to understand how this scales
with photon-number gain $G$ near the bifurcation; this can be done without explicitly Keldysh-ordering operators.  
The leading-order-in-$\bar{n}$ contributions to the fourth cumulant will come from terms of the form
\begin{eqnarray}
	\label{eq:kurtosis1}
	\langle\langle \hat{m}^4 \rangle\rangle_{\rm DPA} & \sim &
		\left( \prod_{j=1}^4 \int_0^t dt_j  \right) \;
		\left \langle 
			(\sqrt{\bar{n}}\hddag(t_1))\cdot(\sqrt{\bar{n}} \hd(t_2))\cdot( \hddag(t_3) \hd(t_3)) \cdot 
				(\hddag(t_4) \hd(t_4)) 
			\right \rangle, \nonumber \\
 	& \sim & 
		\bar{n} \left( \prod_{j=1}^4 \int_0^t dt_j  \right) \; 
			\left \langle \hddag(t_1) \hd(t_3) \right \rangle  \cdot  
			\left \langle\hd(t_2) \hddag(t_4) \right \rangle  \cdot  
			\left \langle \hddag(t_3) \hd(t_4) \right \rangle+...,\nonumber \\
\end{eqnarray}
where in the last line we have applied Wick's theorem, writing only one of the possible pairings explicitly.  In the long-time limit,
only the zero-frequency behaviour of the two-point correlations above will be important.  One finds that near the bifurcation, these susceptibilities scale as \cite{Laflamme2011}:
\begin{equation}
	\int dt \left \langle \hddag(t) \hd(0) \right \rangle \sim 
	\int dt \left \langle \hd(t) \hddag(0) \right \rangle \sim \frac{G}{\kappa},
\end{equation}
where we have dropped terms lower order in $G$. This implies that for a true DPA near bifurcation, 
\begin{eqnarray}
\label{eq:kurtosis2}
\langle\langle \hat{m}^4 \rangle\rangle_{\rm DPA} & \sim &\frac{ \bar{n}G^3 t}{\kappa^3}.
\end{eqnarray}

This does not agree with the full calculation presented in the text, which finds that the kurtosis for the intracavity photon number fluctuations
scales like $G^4$ near the bifurcation (c.f.~Eq.~(8) and (7) in the main text).  The discrepancy is a 
direct result of neglect of the cubic and quartic terms in $\hd$ in the full Hamiltonian of Eq.~(\ref{eq:appendix_Hamiltonian}).  One can obtain the leading $G^4$ behaviour of the kurtosis (in agreement with the results of the main text) by treating these terms perturbatively.  This explicitly demonstrates that the phase-space approach of the main text goes well beyond a simple linearization of the cavity Hamiltonian.

%In contrast to this approach, which neglects all higher-order-in-$\hd$ terms completely, these additional terms can be accounted for when treated as a perturbation to the otherwise quadratic Hamiltonian. Including these terms to first order modifies the leading order term in Eq.~(\ref{eq:kurtosis1}) to be
%\begin{eqnarray}
%\label{eq:m4_terms}
%\langle\langle \hat{m}^4 \rangle\rangle_{\rm DPA} &\sim& \int_0^t dt_1 dt_2 dt_3dt_4dt_5 \;  \langle (\sqrt{\bar{n}}\hddag(t_1))\cdot( \sqrt{\bar{n}}\hd(t_2))\cdot( \sqrt{\bar{n}}\hddag(t_3))\cdot(\sqrt{\bar{n}} \hd(t_4) )\cdot( \Lambda\hddag\hddag\hd\hd(t_5) )\rangle  \nonumber \\
%&\sim& \frac{\bar{n}G^4 t}{\kappa^3}.
%\end{eqnarray}
%In Eq.~(\ref{eq:m4_terms}) we use the result from [12] that in the large $G$ limit $\bar{n} \sim \kappa/\Lambda$, and thus $\Lambda \sim \kappa/\bar{n}$ . Including these additional terms describes the higher scaling of the fourth moment, thus confirming that treating the number fluctuations within a linear theory does not fully describe the results. While the approach used to derive the dephasing rate (ie. solving the equations of motion of the Wigner function parameters in the previous section) given in Eq.~(7) is not perturbative, it does capture this scaling correctly.

\section{Measurement Rate beyond Linear Response}

During a weak measurement information about the system is only acquired slowly, implying that the output homodyne current must be integrated over a finite period of time in order to resolve the qubit state. If the measurement begins at $t=0$ then the time-integrated homodyne intensity, $\hat{s}$, is given by
\begin{equation}
\hat{s}(t)=\int_0^t dt' \hat{I}(t') \sim \int_0^t dt'(\hd_{\rm out}(t') e^{-i\phi}+\hddag_{\rm out}(t') e^{i\phi} ),
\end{equation} 
where $\hd_{\rm out} = \hd_{\rm in} +\sqrt{\kappa}\hd$ is the cavity output field, as defined by input-output theory [4].  Each qubit
eigenstate  $\sigma = \uparrow, \downarrow$ will lead to a different probability distribution for $s$; we denote these two distributions 
$ p_{\sigma}(s)$. The measurement rate characterizes how quickly these two distributions become distinguishable (i.e.~how quickly can we resolve the qubit state from the homodyne current).  As is standard for weak continuous measurements 
\cite{Clerk2010}, we define the measurement rate $\Gamma_{\rm meas}$ by the long-time decay of the overlap of $p_{\uparrow}(s)$ and $p_{\downarrow}(s)$
\begin{equation}
	\label{eq:measrate_def}
	\Gamma_{\rm meas} = - \lim_{t\rightarrow \infty} 
		\frac{  \ln \left[ \int ds\;\sqrt{ p_\uparrow(s)p_\downarrow(s) }\right] }{t}.
\end{equation} 
Further, in the long-time limit (i.e.~times longer than any internal detector timescale), the central limit theorem applies, 
and the distributions $p_\sigma(s)$ will be Gaussian.  Each distribution will thus be fully characterized by the corresponding mean and variance of 
$s$.  These are in turn related to the average homodyne current $\langle \hat{I} \rangle_{\sigma}$ 
and zero-frequency homodyne current noise
\begin{equation}
	S_{II,\sigma}[0] \equiv \int_{-\infty}^{\infty} dt \, \langle \hat{I}(t) \hat{I}(0) \rangle_{\sigma}
\end{equation}
associated with each qubit state.  Using these Gaussian forms and directly integrating Eq.~(\ref{eq:measrate_def}), we obtain the result quoted in Eq.~(9) of the main text.  Note that in the limit of small $\lambda$, Eq.~(9) reduces to the standard linear-response expression for the measurement rate in the weak-coupling limit \cite{Clerk2010}.  The required mean homodyne current and  homodyne current noise in Eq.~(9) can 
be calculated directly using the linearized Hamiltonians $\hH_\sigma$ in Eq.~(3) and input-output theory, following the approach of Ref. \cite{Laflamme2011}.  

While the one can always define a measurement rate in the above manner, it is only a useful quantity when the measurement is slow compared to detector timescales, i.e. $\Gamma_{\rm meas} \cdot t_{\rm detector} \leq 1$.  For our nonlinear cavity detector, the slowest internal
detector timescale is associated with the narrow bandwidth of parametric amplification, and is given by $t_{\rm slow} \sim \sqrt{G} / \kappa$ \cite{Laflamme2011}.  

\section{Efficiency Ratio Beyond Linear Response}
In the case where the coupling between the qubit and the cavity is weak enough that linear response is valid, there exists a fundamental quantum limit on the efficiency of quantum non-demolition (QND) qubit detection. This limit states that the best one can do is measure the qubit as quickly as one dephases it \cite{Clerk2010}, i.e.:
\begin{equation}
	\chi \equiv \frac{\Gamma_{\rm meas}}{\Gamma_\varphi} \leq 1.
	\label{eq:QL}
\end{equation}
While this constraint is most easily derived in the limit of a weak qubit-detector coupling \cite{Clerk2010}, it can be generalized to the stronger couplings we are interested in.  In the general QND case, one can derive a lower bound on the the dephasing rate $\Gamma_{\varphi}$ by considering the most ideal situation, where the backaction dephasing is completely due to the process of information gain.  The ideal case corresponds to the two following requirements:
\begin{enumerate}
	\item If at $t=0$ when the detector-qubit coupled is switched on the qubit is in its eigenstate $| \sigma \rangle$
	 ($\sigma = \uparrow, \downarrow$), then at time $t$
	the detector will be in the pure state $| D_{\sigma}(t) \rangle$.
	\item The states $| D_{\uparrow}(t) \rangle, | D_{\downarrow}(t) \rangle$ are completely determined by the corresponding 
	probability distributions $p_{\sigma}(s)$ of the measured detector quantity (i.e.~in our case, the integrated homodyne current).
\end{enumerate}

The first requirement implies that if the qubit is initially in a superposition of its eigenstates 
(i.e. $a | \uparrow \rangle + b |\downarrow \rangle$), then at at time $t$ it would be entangled with the detector.  The detector-qubit
system would be described by the state
\begin{equation}
|\Psi(t) \rangle =   a |\uparrow\rangle\otimes|D_\uparrow\rangle
	+ b |\downarrow\rangle\otimes|D_\downarrow\rangle.
\end{equation}
Using the definition of Eq.~(\ref{eq:DephRateDefn}), the long-time dephasing rate would then be given by the overlap of the two detector pointer states:
\begin{equation}
	\Gamma_{\varphi} = -\lim_{t \rightarrow \infty} \ln \frac{ | \langle D_\uparrow | D_\downarrow \rangle | }{t}
	\label{eq:OverlapDephasing}
%
%\langle \downarrow | \hat{\rho}|\uparrow\rangle = \frac{e^{i\phi}}{2}\langle D_\uparrow | D_\downarrow \rangle\sim e^{-\Gamma_{\varphi}t},
\end{equation}

The second requirement above further implies that if $| s \rangle$ represents a state with definite value of the detector output (i.e. integrated homodyne current), then the states $|D_\sigma \rangle$ can be written:
\begin{equation}
	\label{eq:detector_state}
	|D_{\sigma}\rangle =\int ds\; \sqrt{p_{\sigma}(s)}|s\rangle.
\end{equation}

Using the definition of the measurement rate (Eq.~(\ref{eq:measrate_def})) and the expression for the dephasing rate above (Eq.~(\ref{eq:OverlapDephasing})), we thus find in this most ideal case:
\begin{equation}
\Gamma_\varphi = \Gamma_{\rm meas},
\end{equation}
i.e.~the measurement rate and dephasing rate coincide.  This represents a lower bound on the dephasing.  In the more general case, the overlap of the the detector pointer states will be {\it less} than that implied by Eq.~(\ref{eq:detector_state}), as the qubit will also become entangled with degrees of freedom in the detector not directly related to the observed quantity $s$.  Explicitly, in the more general case we would have:
\begin{equation}
	\label{eq:detector_state_full}
	 |D_{\sigma}\rangle = \left( \int ds\; \sqrt{p_{\sigma}(s)}|s\rangle \right) \otimes |E_{\sigma} \rangle.
\end{equation}
where the states $|E_{\sigma} \rangle$ describe the additional detector degrees of freedom.  If $| \langle E_{\uparrow} | E_{\downarrow} \rangle | < 1$, then there is ``wasted information" information in the detector (i.e. information on the qubit state in the unobserved degrees of freedom described by $|E_{\sigma}\rangle$), and the dephasing rate will be necessarily less than the ideal value $\Gamma_{\rm meas}$.  We have thus proved the quantum limit inequality Eq.~(\ref{eq:QL}) without invoking a small detector-qubit coupling. 
%
%In the more general case Eq.~(\ref{eq:detector_state}) will not be true, rather there will be extra external degrees of freedom which determine the detector state and therefore contribute to additional sources of dephasing. Thus, in an arbitrary setup the dephasing rate will satisfy the limit general beyond linear response 
%\begin{equation}
%\Gamma_\varphi \geq \Gamma_{\rm meas}.
%\end{equation}
% \twocolumn
%\end{singlecolumn}

%%%%%%%%%%%%%%%%%